\documentclass[a4paper,twocolumn,aps]{revtex4}
\usepackage[T1]{fontenc}
\usepackage{graphics}

\makeatletter

\providecommand{\LyX}{L\kern-.1667em\lower.25em\hbox{Y}\kern-.125emX\@}

\usepackage{amssymb}

\usepackage{graphicx}
\usepackage{dcolumn}
\usepackage{amsmath}
\makeatother

\begin{document}

\title{Spin susceptibility of the superfluid \protect\( ^{3}\protect \)He-B in aerogel }

\author{V. P. Mineev }

\address{D\'epartament de Recherche Fondamentale sur la Mati\`ere Condens\'ee, Commisariat
\`a l'\'Energie Atomique, Grenoble 38054, France}

\author{P. L. Krotkov}

\address{L. D. Landau Institute for Theoretical Physics, Russian Academy of Sciences,
117334 Moscow, Russia}

\date{\today}

\begin{abstract}
The temperature dependence of paramagnetic susceptibility of the superfluid
\( ^{3} \)He-B in aerogel is found. Calculations have been performed for an
arbitrary phase shift of \( s \)-wave scattering in the framework of BCS weak
coupling theory and the simplest model of aerogel as an aggregate of homogeneously
distributed ordinary impurities. Both limiting cases of the Born and unitary
scattering can be easily obtained from the general result. The existence of
gapless superfluidity starting at the critical impurity concentration depending
on the value of the scattering phase has been demonstrated. While larger than
in the bulk liquid the calculated susceptibility of the B-phase in aerogel proves
to be conspicuously smaller than that determined experimentally in the high
pressure region.

\end{abstract}
\maketitle
\newcommand{\f}[1]{{\mbox{\boldmath$#1$}}}

\newcommand{\tr}{\mathop{\rm tr_{\sigma}}\nolimits}

\newcommand{\sgn}{\mathop{\rm sgn}\nolimits}

\newcommand{\Tr}{\mathop{\rm tr_{\tau}}\nolimits} 

\newcommand{\TS}{\textstyle} 

\newcommand{\DS}{\displaystyle} 

Superfluid \( ^{3} \)He is an ideal object for a study of physical properties
of a system with non-trivial pairing. In particular, it is interesting to investigate
how a superfluid state could be influenced by the presence of impurities. Direct
contamination of \( ^{3} \)He with any atomic impurities is impossible. Last
years instead of this experimentalists use liquid \( ^{3} \)He to fill up aerogel
which is a matrix of randomly arranged silica filaments of nanometer diameter.
Commonly used aerogels occupy about 2\% of space volume, the rest being taken
with liquid helium.

Experiments \cite{porto95}, \cite{sprague95}, \cite{sprague96}, \cite{matsumoto97},
\cite{lawes00} established decrease of the superfluid transition temperature
and the density of the superfluid component of \( ^{3} \)He in aerogel. These
effects are related to the scattering of quasiparticles from surface, which
suppresses Cooper pairing in the \( p \)-state. Corresponding theoretical treatments
have been proposed in papers \cite{Thuneberg98}, \cite{Rainer98}, \cite{Hanninen98},
\cite{Hanninen00}, \cite{Higashitani99}.

NMR measurements \cite{sprague95} pointed on the existence of the A-phase of
\( ^{3} \)He in aerogel in relatively high magnetic fields as well as of the
B-phase \cite{alles99} in lower fields. Recently, evidence of the phase transition
between the A- and B-phases has been demonstrated \cite{Osheroff00}, \cite{brussard01},
\cite{gervais01}.

The A-phase as an equal spin pairing state possesses practically the same spin
susceptibility as the normal phase. This property remains intact in the A-phase
in aerogel. On the other hand, in the B-phase where all the three spin states
of the Cooper pairs with the spin projections \( S_{z}=0,\pm 1 \) are equally
populated, spin susceptibility is partly suppressed compared to its normal state
value. This suppression in aerogel was experimentally found \cite{Osheroff00}
to constitute about 45\% of the value in the bulk \( ^{3} \)He-B.

The purpose of the present article is to calculate spin susceptibility of the
\( ^{3} \)He-B in aerogel. We shall work in the framework of BCS weak coupling
theory, aerogel being approximated as homogeneously distributed ordinary impurities.
At high pressures this model gives values of \( \Delta  \) and \( \rho _{s} \)
roughly by a factor of two larger \cite{Thuneberg98}, \cite{Hanninen98} than
those observed experimentally. At the same time, as it was discussed in paper
\cite{porto99}, at low pressures, when the coherence length is larger than
the average pore size, the homogeneous scattering model is more likely to be
justified. 

A similar conclusion follows from the present derivation. The calculated susceptibility
is less suppressed compared to the bulk dependence. This suppression is less
pronounced for larger values of the scattering phase but still even in the unitary
limit noticeably larger than the measured suppression of the susceptibility
at high pressures \cite{sprague96}, \cite{Osheroff00}.

The calculations presented in the paper have been done for an arbitrary value
\( \delta _{0} \) of the phase shift of \( s \)-wave scattering. Both extremes:
the Born approximation and the unitary limit can be easily obtained from the
general result. 

Corresponding theory of changes of paramagnetic susceptibility in ordinary superconductors
caused by impurities with ordinary, paramagnetic and spin-orbital type of scattering
has been developed in the Born approximation in paper \cite{Abrikosov62}. The
present theory is not a simple generalization of the cited paper. The point
is that usually to come out of the Born approximation in the theory of dirty
alloys it is sufficient to substitute the Born scattering amplitude by its exact
counterpart \cite{Abrikosov63}. This substitution does not lead to the correct
answer for a dirty metal in a magnetic field. The method developed in the present
paper does not rely on the substitution.

The paper is organized as follows. After the introduction of the basics of the
Abrikosov-Gor'kov theory in the first section, we write down its solution for
an arbitrary impurity scattering phase without magnetic field in the second
section, where we also establish the region of the existence of gapless superfluidity.
Then by finding corrections in leading order in magnetic field we arrive at
an expression for the susceptibility. In the last section we briefly discuss
conclusions.

\section{Equations}

The Abrikosov-Gor'kov equations of an electrically neutral impure superfluid
in an external magnetic field \( \mathbf{B} \) are (see, e.g. \cite{Mineev99})
\begin{equation}
\label{AG}
\left( i\omega _{n}-\widehat{H}(\mathbf{k})-\widehat{\Sigma }(\omega _{n})\right) \widehat{G}(\mathbf{k},\omega _{n})=\widehat{1},
\end{equation}
 where \( \omega _{n}=\pi T(2n+1) \) is the Matsubara frequency. Green function
\( \widehat{G}(\mathbf{k},\omega _{n}) \) is a 2x2 matrix in particle-hole
space 
\begin{equation}
\label{G}
\widehat{G}(\mathbf{k},\omega _{n})=\left( \begin{array}{cc}
G_{\alpha \beta }(\mathbf{k},\omega _{n}) & F_{\alpha \beta }(\mathbf{k},\omega _{n})\\
F_{\alpha \beta }^{+}(\mathbf{k},\omega _{n}) & \overline{G}_{\alpha \beta }(\mathbf{k},\omega _{n})
\end{array}\right) ,
\end{equation}
 made up of the normal \( G_{\alpha \beta }(\mathbf{k},\omega _{n}) \) and
anomalous \( F_{\alpha \beta }(\mathbf{k},\omega _{n}) \) Green functions which
are in their turns 2x2 matrices in spin space. Here 
\begin{equation}
\overline{G}_{\alpha \beta }(\mathbf{k},\omega _{n})=-G_{\alpha \beta }^{\mathrm{T}}(-\mathbf{k},-\omega _{n})
\end{equation}
 and the superscript T implies transposition. 

The Hamiltonian 
\begin{equation}
\label{H}
\widehat{H}(\mathbf{k})=\left( \begin{array}{cc}
H_{0\alpha \beta }(\mathbf{k}) & \Delta _{\hat{k}\alpha \beta }\\
\Delta _{\hat{k}\alpha \beta }^{+} & -H_{0\alpha \beta }^{\mathrm{T}}(-\mathbf{k})
\end{array}\right) 
\end{equation}
 consists of the one-particle part 
\begin{equation}
H_{0\alpha \beta }(\mathbf{k})=\xi _{\mathbf{k}}\delta _{\alpha \beta }-\mu \f \sigma _{\alpha \beta }\mathbf{B}
\end{equation}
 including kinetic 
\begin{equation}
\xi _{\mathbf{k}}\equiv \xi _{|\mathbf{k}|}=k^{2}/2m^{\ast }-\epsilon _{F}
\end{equation}
 and Zeeman energy, and of the pairing interaction \( V_{\alpha \beta ,\lambda \mu }(\mathbf{k},\mathbf{k}^{\prime }) \)
via the order parameter \( \Delta _{\hat{k}\alpha \beta } \), set by the self-consistency
equation 
\begin{equation}
\label{D}
\Delta _{\hat{k}\alpha \beta }=-T\sum _{n}\sum _{\mathbf{k}^{\prime }}V_{\beta \alpha ,\lambda \mu }(\mathbf{k},\mathbf{k}^{\prime })F_{\mu \lambda }(\omega _{n},\mathbf{k}^{\prime }).
\end{equation}
 \( \mathbf{B} \) is the magnetic flux, \( \f \sigma _{\alpha \beta } \) is
a vector of the three Pauli matrices in spin space.

The impurity scattering self-energy part comprises normal \( \Sigma _{1} \)
and anomalous \( \Sigma _{2} \) parts 
\begin{equation}
\widehat{\Sigma }=\left( \begin{array}{cc}
\Sigma _{1} & \Sigma _{2}\\
\Sigma _{2}^{+} & \overline{\Sigma }_{1}
\end{array}\right) ,
\end{equation}
 and in the clean limit of small impurity concentrations \( n_{\mathrm{imp}} \),
when one can neglect the interference of the scattering on different impurities,
is just \( n_{\mathrm{imp}} \) times the contribution of a single impurity
\begin{equation}
\label{Sigma}
\widehat{\Sigma }=n_{\mathrm{imp}}\widehat{T},
\end{equation}
 where \( \widehat{T} \) is the scattering amplitude. It is related to the
scattering potential \( \widehat{U}_{\mathbf{kk}^{\prime }} \) by the Lippmann-Schwinger
equation 
\begin{equation}
\label{Lippmann-Schwinger}
\widehat{T}_{\mathbf{kk}^{\prime }}=\widehat{U}_{\mathbf{kk}^{\prime }}+\sum _{\mathbf{q}}\widehat{U}_{\mathbf{kq}}\widehat{G}_{\mathbf{q}}\widehat{T}_{\mathbf{qk}^{\prime }}.
\end{equation}
 In the simplest case of a short-range impurity potential \( U(\mathbf{r})=u\delta (\mathbf{r}) \)
its Fourier component is momentum-independent \( U_{\mathbf{k}-\mathbf{k}^{\prime }}=u \),
so that 
\begin{equation}
\label{U}
\widehat{U}_{\mathbf{kk}^{\prime }}\equiv \left( \begin{array}{cc}
U_{\mathbf{k}-\mathbf{k}^{\prime }}\delta _{\alpha \beta } & 0\\
0 & -U_{\mathbf{k}^{\prime }-\mathbf{k}}\delta _{\alpha \beta }
\end{array}\right) =u\delta _{\alpha \beta }\widehat{\tau }_{3},
\end{equation}
 where \( \widehat{\f \tau } \) is a vector of the three Pauli matrices in
particle-hole space. Parameterized in terms of the scattering phase \( \delta _{0} \)
the impurity potential is 
\begin{equation}
u=-\tan \delta _{0}/\pi N_{0}.
\end{equation}

Eq. (\ref{Lippmann-Schwinger}) is then easily solved, and the scattering amplitude
turns out to be independent of momenta 
\begin{equation}
\label{T}
\widehat{T}_{\mathbf{kk}^{\prime }}(\omega _{n})\equiv \widehat{T}(\omega _{n})=\widehat{U}\left( \widehat{1}-\widehat{U}\sum _{\mathbf{k}}\widehat{G}(\mathbf{k},\omega _{n})\right) ^{-1}.
\end{equation}

Equations (\ref{AG}), (\ref{H}), (\ref{Sigma}), (\ref{T}) form a closed
system of equations on the Green function (\ref{G}).

The pairing interaction \( V_{\beta \alpha ,\lambda \mu }(\mathbf{k},\mathbf{k}^{\prime }) \)
is usually reckoned non-zero only in a thin \( \sim \epsilon _{1}\ll \epsilon _{F} \)
layer near the Fermi surface. In \( ^{3} \)He spin-orbital interaction is weak
and the Cooper pairs are formed in the spin-triplet, orbital \( p \)-wave state.
So the pairing interaction can be factorized 
\begin{equation}
\label{pairing-pot}
V_{\beta \alpha ,\lambda \mu }(\mathbf{k},\mathbf{k}^{\prime })=-\frac{3}{2}V_{1}\hat{k}\hat{k}^{\prime }\mathbf{g}_{\beta \alpha }\mathbf{g}_{\lambda \mu }^{+},
\end{equation}
where \( V_{1} \) is the constant of the \( p \)-wave pairing attraction and
\begin{equation}
\mathbf{g}_{\alpha \beta }=i(\f \sigma \sigma _{y})_{\alpha \beta }
\end{equation}
 is a vector of the three basis symmetric matrices. 

The order parameter 
\begin{equation}
\label{dk}
\Delta _{\hat{k}\alpha \beta }=\mathbf{d}_{\hat{k}}\mathbf{g}_{\alpha \beta },
\end{equation}
 where \( d_{\mu \hat{k}}=A_{\mu i}\hat{k}_{i} \) is the order-parameter vector. 

Substituting (\ref{pairing-pot}) and (\ref{dk}) into (\ref{D}) yields 
\begin{equation}
\label{dk2}
\mathbf{d}_{\hat{k}}=\frac{3}{2}V_{1}T\sum _{n}\sum _{\mathbf{k}^{\prime }}(\hat{k}\hat{k}^{\prime })\tr [\mathbf{g}^{+}F(\omega _{n},\mathbf{k}^{\prime })].
\end{equation}

As one can see 
\begin{equation}
\mathbf{d}_{\hat{k}}^{\mathstrut }\mathbf{d}_{\hat{k}}^{\mathstrut \ast }=\frac{1}{2}\tr [\Delta _{\hat{k}}^{+}\Delta _{\hat{k}}^{\mathstrut }],
\end{equation}
where \( \tr  \) denotes a trace over spin indices. For unitary phases, among
which are the A- and B-phases actually being realized in pure \( ^{3} \)He,
(\( \Delta _{\hat{k}}^{+}\Delta ^{\mathstrut }_{\hat{k}})_{\alpha \beta }\propto \delta _{\alpha \beta } \)
and so it is convenient to introduce the notation \( \Delta _{\hat{k}}^{\mathstrut 2}=\mathbf{d}_{\hat{k}}^{\mathstrut }\mathbf{d}_{\hat{k}}^{\mathstrut \ast } \).
We will consider only the unitary phases.

In the A-phase 
\begin{equation}
\mathbf{d}_{\hat{k}}=\sqrt{\frac{3}{2}}\Delta \hat{d}[(\hat{\Delta }^{\prime }+i\hat{\Delta }^{\prime \prime })\hat{k}],
\end{equation}
where \( \hat{d} \), \( \hat{\Delta }^{\prime } \), and \( \hat{\Delta }^{\prime \prime } \)
are three unit vectors, \( \hat{\Delta }^{\prime } \) and \( \hat{\Delta }^{\prime \prime } \)
are mutually orthogonal.

In the B-phase 
\begin{equation}
\mathbf{d}_{\hat{k}}=\Delta \overleftrightarrow {R}\hat{k}e^{i\varphi },
\end{equation}
where \( \overleftrightarrow {R} \) is the matrix of rotation on an arbitrary
angle.

In both cases, the scalar \( \Delta  \) is introduced so as to fulfill the
normalization \( \langle \Delta _{\hat{k}}^{\mathstrut 2}\rangle _{\hat{k}}^{\mathstrut }=\langle \mathbf{d}_{\hat{k}}^{\mathstrut }\mathbf{d}_{\hat{k}}^{\mathstrut \ast }\rangle _{\hat{k}}^{\mathstrut }=\Delta ^{2} \),
where \( \langle \ldots \rangle _{\hat{k}}^{\mathstrut } \) stands for the
averaging over directions \( \hat{k} \) of momentum.

\section{Zero magnetic field}

In the absence of magnetic field the solution of Eqs. (\ref{AG}), (\ref{H}),
(\ref{Sigma}), (\ref{T}) is (see elsewhere) 
\begin{eqnarray}
\widehat{G}^{(0)} & = & \frac{1}{\widetilde{\omega }_{n}^{2}+\widetilde{\xi }_{\mathbf{k}}^{2}+\Delta _{\hat{k}}^{2}}\label{G0} \\
 & \times  & \left( \begin{array}{cc}
-(i\widetilde{\omega }_{n}+\widetilde{\xi }_{\mathbf{k}})\delta _{\alpha \beta } & \Delta _{\hat{k}\alpha \beta }\\
\Delta _{\hat{k}\alpha \beta }^{+} & -(i\widetilde{\omega }_{n}-\widetilde{\xi }_{\mathbf{k}})\delta _{\alpha \beta }
\end{array}\right) .\nonumber 
\end{eqnarray}

The corresponding sum over momenta in the expression (\ref{T}) for the scattering
amplitude is 
\[
\sum _{\mathbf{k}}\widehat{G}^{(0)}(\mathbf{k},\omega _{n})=g(\omega _{n})\widehat{1}\delta _{\alpha \beta },\]
where we denoted 
\begin{eqnarray}
g(\omega _{n}) & = & \sum _{\mathbf{k}}G^{(0)}=-N_{0}\! \! \int \! \! \left\langle \frac{i\widetilde{\omega }_{n}+\widetilde{\xi }}{\widetilde{\omega }_{n}^{2}+\widetilde{\xi }^{2}+\Delta _{\hat{k}}^{2}}\right\rangle _{\! \! \hat{k}}d\widetilde{\xi }\nonumber \\
 & = & -i\pi N_{0}\widetilde{\omega }_{n}\left\langle \frac{1}{\sqrt{\widetilde{\omega }_{n}^{2}+\Delta _{\hat{k}}^{2}}}\right\rangle _{\! \! \hat{k}}.\label{SumkG0} 
\end{eqnarray}
Here \( N_{0}=m^{*}k_{F}/2\pi ^{2}\hbar ^{2} \) is the density of states on
the Fermi surface.

So the scattering self-energy (\ref{Sigma}) equals
\begin{equation}
\label{Sigma0}
\widehat{\Sigma }^{(0)}=\frac{n_{\mathrm{imp}}u}{1-u^{2}g^{2}}\left[ \widehat{\tau }_{3}+ug\widehat{1}\right] \delta _{\alpha \beta }.
\end{equation}

In the above formulae both the Matsubara frequency \( \omega _{n}=\pi T(2n+1) \)
and the kinetic energy are renormalized 
\begin{eqnarray}
i\widetilde{\omega }_{n} & = & i\omega _{n}+\frac{1}{2}\Tr [\widehat{\Sigma }^{(0)}(\omega _{n})],\label{omegaTilde} \\
\widetilde{\xi }_{\mathbf{k}} & = & \xi _{\mathbf{k}}-\frac{1}{2}\Tr [\widehat{\tau }_{3}\widehat{\Sigma }^{(0)}(\omega _{n})].\label{xik} 
\end{eqnarray}
Here \( \Tr  \) is the trace in particle-hole space. 

It should be noted that the term proportional to \( \widehat{\tau }_{3} \)
in \( \widehat{\Sigma }=\widehat{\Sigma }^{(0)}+\widehat{\Sigma }^{(1)}+\ldots  \)
must be omitted. This term just produces a shift in the chemical potential (renormalization
of \( \xi _{\mathbf{k}} \) ) from introducing the impurities and disappears
in the assumption of particle-hole symmetry after integration over \( \xi  \)
in the subsequent calculations. 

Substituting (\ref{SumkG0}) and (\ref{Sigma0}) into (\ref{omegaTilde}) yields
the self-consistency equation on the renormalized Matsubara frequency \( \widetilde{\omega }_{n} \)
\begin{equation}
\label{omegaTilde2}
\widetilde{\omega }_{n}=\omega _{n}+\Gamma \left\langle \frac{\widetilde{\omega }_{n}\sqrt{\widetilde{\omega }_{n}^{2}+\Delta _{\hat{k}}^{2}}}{\widetilde{\omega }_{n}^{2}+\cos ^{2}\delta _{0}\Delta _{\hat{k}}^{2}}\right\rangle _{\! \! \hat{k}},
\end{equation}
 where \( \Gamma  \) is the scattering rate 
\begin{equation}
\label{Gamma}
\Gamma =\frac{n_{\mathrm{imp}}}{\pi N_{0}}\sin ^{2}\delta _{0}.
\end{equation}
 In the Born limit (\( \delta _{0}\rightarrow 0 \)) it tends to the conventional
half of the inverse free flight time 
\begin{equation}
\Gamma \rightarrow \frac{n_{\mathrm{imp}}}{\pi N_{0}}\delta _{0}^{2}=n_{\mathrm{imp}}\pi N_{0}u^{2}=\frac{1}{2\tau }.
\end{equation}

In the B-phase \( \Delta _{\hat{k}}^{2}\equiv \Delta ^{2} \) does not depend
on \( \hat{k} \), one can thus omit angular brackets of the \( \hat{k} \)-averaging
in (\ref{omegaTilde2}). Using the overt form (\ref{G0}), the self-consistency
Eq. (\ref{dk2}) taking the trace over spin indices and averaging over \( \hat{k} \)
reduces to 
\begin{equation}
\label{1/NV}
\frac{1}{N_{0}V_{1}}=T\sum _{n}\int \frac{d\xi }{\widetilde{\omega }_{n}^{2}+\xi ^{2}+\Delta ^{2}}=\pi T\sum _{n}\frac{1}{\sqrt{\widetilde{\omega }_{n}^{2}+\Delta ^{2}}}.
\end{equation}

This equation determines the temperature and impurity concentration behavior
of the order parameter \( \Delta  \). To obtain an expression of value one
should exclude the unobservable pairing constant \( V_{1} \) off the left-hand
side of Eq. (\ref{1/NV}). Using the standard procedure \cite{Mineev99}, one
obtains the self-consistency equation on \( \Delta (T,\Gamma ) \)
\begin{equation}
\label{logTTc0}
\log \frac{T}{T_{c0}}=\pi T\sum _{n}\left( \frac{1}{\sqrt{\widetilde{\omega }_{n}^{2}+\Delta ^{2}}}-\frac{1}{|\omega _{n}|}\right) .
\end{equation}
 The asymptotics of \( \Delta (T,\Gamma ) \) at \( T=0,T_{c} \) are found
in the Appendix.

Expanding (\ref{omegaTilde2}) in degrees of \( \Delta ^{2} \) gives \( \widetilde{\omega }_{n}\approx \omega _{n}+\Gamma \sgn \omega _{n}+O(\Delta ^{2}) \).
When inserted into (\ref{logTTc0}) it yields in zeroth order in \( \Delta ^{2} \)
the Abrikosov-Gor'kov \cite{Abrikosov61} implicit expression for the dependence
of the critical temperature \( T_{c} \) on the impurity concentration 
\begin{equation}
\log \frac{T_{c0}}{T_{c}}=\psi \left( \frac{\Gamma }{2\pi T_{c}}+\frac{1}{2}\right) -\psi \left( \frac{1}{2}\right) ,
\end{equation}
 where \( \psi  \) is the digamma function. Note however that \( \Gamma  \)
corresponds to different impurity concentrations for various scattering phases
(see (\ref{Gamma})). \( T_{c}(\Gamma ) \) vanishes at the critical concentration
\begin{equation}
\Gamma _{c}=\TS \frac{1}{2}\Delta _{00},
\end{equation}
where 
\begin{equation}
\label{Delta00}
\Delta _{00}=\pi T_{c0}e^{-\gamma }
\end{equation}
 is the order parameter at zero temperature for a pure superfluid, \( \gamma \approx 0.5772 \)
is Euler's constant.

\begin{figure}[!t]
{\par\centering \resizebox*{1\columnwidth}{!}{\includegraphics{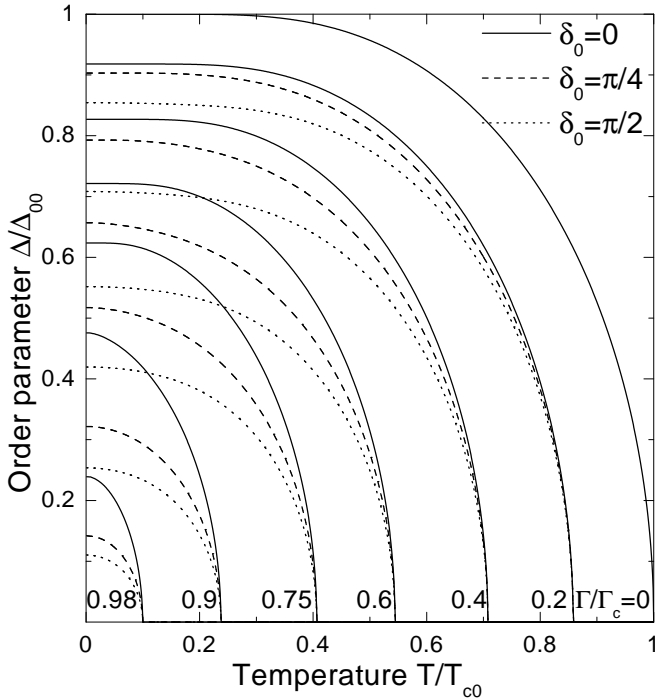}} \par}

\caption{\label{Fig-DeltaOnT} Order parameter \protect\( \Delta \protect \) versus
temperature \protect\( T\protect \) for different scattering phases \protect\( \delta _{0}\protect \)
and impurity concentrations obtained numerically. Solid lines correspond to
the Born limit \protect\( \delta _{0}=0\protect \), dashed lines to \protect\( \delta _{0}=\frac{\pi }{4}\protect \)
and dotted lines to the unitary scattering \protect\( \delta _{0}=\frac{\pi }{2}\protect \).
In order of the decrease of the superfluid transition temperature \protect\( T_{c}\protect \)
the triples of the curves correspond to \protect\( \Gamma /\Gamma _{c}=\protect \)0,
0.2, 0.4, 0.6, 0.75, 0.9, 0.98. The first curve in the pure case \protect\( \Gamma =0\protect \)
is common for all phases.}
\end{figure}

Fig. \ref{Fig-DeltaOnT} shows temperature dependences of the order parameter
for 7 impurity concentrations and for three scattering phases: \( \delta _{0}=0,\frac{\pi }{4} \),
and \( \frac{\pi }{2} \) obtained numerically from (\ref{logTTc0}). Non-shown
plots for intermediate \( 0<\delta _{0}<\frac{\pi }{2} \) all sit in the interior
of the belt between curves for \( \delta _{0}=0 \) (Born limit) and \( \delta _{0}=\frac{\pi }{2} \)
(unitary limit).

\begin{figure}[!t]
{\par\centering \resizebox*{1\columnwidth}{!}{\includegraphics{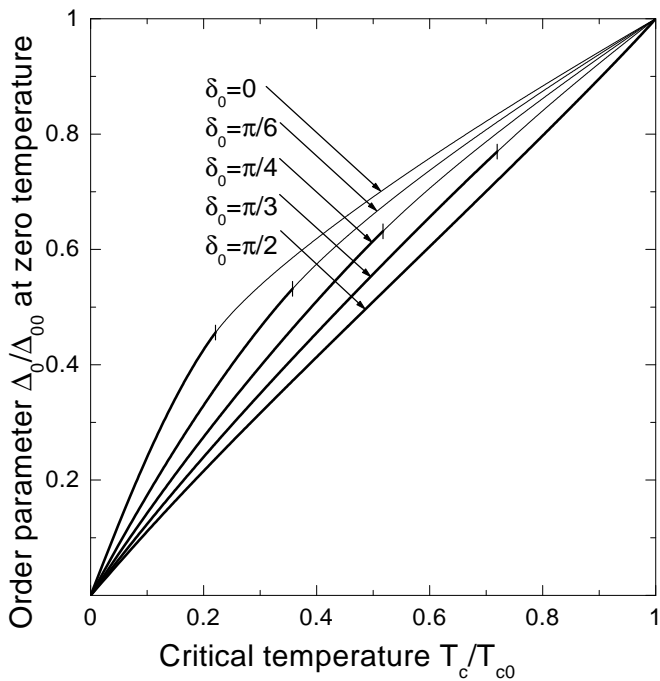}} \par}

\caption{\label{Fig-Delta_0} Order parameter \protect\( \Delta _{0}\protect \) at
zero temperature as a function of the suppression \protect\( T_{c}/T_{c0}\protect \)
of the critical temperature for several values of the scattering phase \protect\( \delta _{0}\protect \).
For each \protect\( \delta _{0}\protect \) the value of \protect\( T_{c}/T_{c0}\protect \)
is pointed, the suppression below which makes the superfluid gapless in the
whole range \protect\( 0<T<T_{c}\protect \). In the unitary limit (\protect\( \delta _{0}\to \frac{\pi }{2}\protect \))
the superfluid is gapless for whatever \protect\( T_{c}/T_{c0}\protect \).
Bold parts on each curve mark the regions of the existence of gapless superfluidity.}
\end{figure}

The order parameter \( \Delta _{0} \) at zero temperature as a function of
the scattering rate \( \Gamma /\Gamma _{c} \) is presented on Fig. \ref{Fig-Delta_0}
for several values of the scattering phase \( \delta _{0} \). The curves for
intermediate values of \( 0<\delta _{0}<\frac{\pi }{2} \) lie inside the strap
between the curves for the Born (\( \delta _{0}\to 0 \)) and unitary (\( \delta _{0}\to \frac{\pi }{2} \))
limits. 

From the Abrikosov-Gor'kov theory of paramagnetic impurities in alloys it is
known that in the Born limit (\( \delta _{0}\to 0 \)) starting at the scattering
rate of 
\begin{equation}
\label{AGGc}
2\Gamma _{c}e^{-\pi /4}\approx 0.91\Gamma _{c}
\end{equation}
 the energy spectrum gap vanishes though the amplitude \( \Delta  \) of the
order parameter remains finite until \( \Gamma _{c} \). Ordinary impurities
in a \( p \)-wave superfluid will give rise to the existence of an analogous
gapless region of impurity concentrations. Indeed, density of states of quasiparticles
at an energy \( \epsilon  \) off the Fermi level is expressed in terms of the
Green function as 
\begin{equation}
N(\epsilon )=-\frac{1}{\pi }\Im \sum _{\mathbf{k}}\left. G(\mathbf{k},\omega _{n})\right| _{i\omega _{n}\rightarrow \epsilon +i0}.
\end{equation}
 Using (\ref{SumkG0}) we get 
\begin{equation}
\label{Ne}
N(\epsilon )=N_{0}\Im \frac{\widetilde{\epsilon }}{\sqrt{\Delta ^{2}-\widetilde{\epsilon }^{2}}},
\end{equation}
where \( \widetilde{\epsilon }=\left. i\widetilde{\omega }_{n}\right| _{i\omega _{n}\rightarrow \epsilon +i0} \).
From (\ref{omegaTilde2}) we obtain an implicit equation on \( \widetilde{\epsilon } \)
for a given \( \epsilon  \):
\begin{equation}
\widetilde{\epsilon }=\epsilon +\Gamma \frac{\widetilde{\epsilon }\sqrt{\Delta ^{2}-\widetilde{\epsilon }^{2}}}{\Delta ^{2}\cos ^{2}\delta _{0}-\widetilde{\epsilon }^{2}}.
\end{equation}
For \( \epsilon =0 \) it gives three solutions in the complex plane: \( \widetilde{\epsilon }_{1}=0 \)
and
\begin{equation}
\label{e23}
\widetilde{\epsilon }^{2}_{2,3}=\frac{2\Delta ^{2}\cos ^{2}\delta _{0}-\Gamma ^{2}-\Gamma \sqrt{\Gamma ^{2}+4\Delta ^{2}\sin ^{2}\delta _{0}}}{2}.
\end{equation}
It can be verified that for all the three solutions \( \Delta ^{2}-\widetilde{\epsilon }^{2} \)
is always (real) positive, and so \( \sqrt{\Delta ^{2}-\widetilde{\epsilon }^{2}} \)
always real. In order for \( N(\epsilon =0) \) to be finite one than sees from
(\ref{Ne}) that \( \widetilde{\epsilon }^{2} \) should be negative. From (\ref{e23})
this gives 
\begin{equation}
\label{GL}
\Gamma >\Delta \cos ^{2}\delta _{0}.
\end{equation}
This inequality means that for a given phase shift \( \delta _{0} \) and a
scattering rate \( \Gamma  \) the superfluid is gapless at temperatures \( T_{\mathrm{gl}}<T<T_{c} \),
where \( T_{\mathrm{gl}} \) is such that \( \Gamma >\Delta (T_{\mathrm{gl}})\cos ^{2}\delta _{0} \).
The superfluid is gapless in the whole temperature range \( 0<T<T_{c} \) of
the existence of superfluidity if inequality (\ref{GL}) is fulfilled already
for \( \Delta _{0}\equiv \Delta (T=0) \). 

Making use of the implicit dependence (\ref{implOnD0}) of \( \Delta _{0} \)
on \( \delta _{0} \), we get for the lower boundary of the onset of gapless
superfluidity in the whole temperature range
\begin{equation}
\Gamma =2\Gamma _{c}\cos ^{2}\delta _{0}e^{-\frac{\pi }{2}\frac{\cos ^{2}\delta _{0}}{1+\cos \delta _{0}}},
\end{equation}
which is a generalization of the Abrikosov-Gor'kov value (\ref{AGGc}) on an
arbitrary scattering phase \( \delta _{0} \). The corresponding value of the
upper boundary of the order parameter at zero temperature is 
\begin{equation}
\Delta _{0}=\Delta _{00}e^{-\frac{\pi }{2}\frac{\cos ^{2}\delta _{0}}{1+\cos \delta _{0}}}
\end{equation}

When the scattering phase is increased (impurity potential becomes ``stronger'')
the boundary value of \( \Gamma  \) decreases, meaning that starting from less
impurity concentration the superfluidity becomes gapless. At last, in the unitary
limit \( \delta _{0}\rightarrow \frac{\pi }{2} \) the boundary value is \( \Gamma =0 \),
and the liquid is gapless for whatever small impurity concentration. 

The points of the onset of gapless superfluidity are marked on the plot of the
dependence of the order parameter at zero temperature \( \Delta _{0}/\Delta _{00} \)
on the suppression \( T_{c}/T_{c0} \) of the superfluid transition temperature
(see Fig. \ref{Fig-Delta_0}). The regions of the existence of gapless superfluidity
are shown on the plots of \( \Delta _{0}(T_{c}) \) with bold lines.

\section{Magnetic susceptibility}

Magnetization \( \mathbf{M} \) can be calculated from the normal Green function
in a magnetic field \( \mathbf{B} \) as
\begin{equation}
\label{M}
\mathbf{M}=\mu T\sum _{n}\sum _{\mathbf{k}}G_{\alpha \beta }\f \sigma _{\beta \alpha }.
\end{equation}
 To find the static susceptibility \( \chi _{\alpha \beta } \) (\( M_{\alpha }=\chi _{\alpha \beta }B_{\beta } \))
it is sufficient to solve Eqs. (\ref{AG}), (\ref{H}), (\ref{Sigma}), (\ref{T})
in leading order in the field. Let the superscript (1) designate the quantities
proportional to \( \mathbf{B} \). From (\ref{Sigma}), (\ref{T}) we get 
\begin{equation}
\label{Sigma1}
\widehat{\Sigma }^{(1)}=\frac{1}{n_{\mathrm{imp}}}\widehat{\Sigma }^{(0)}\widehat{U}(\sum _{\mathbf{k}}\widehat{G}^{(1)})\widehat{U}^{-1}\widehat{\Sigma }^{(0)}.
\end{equation}

One should plug here the expression for the Green function in first order in
the field derived from (\ref{AG}) 
\begin{equation}
\label{Ghat1}
\widehat{G}^{(1)}=\widehat{G}^{(0)}\left( \widehat{H}^{(1)}+\widehat{\Sigma }^{(1)}\right) \widehat{G}^{(0)}.
\end{equation}

We are now going to show that both the anomalous self-energy \( \Sigma _{2} \)
remains zero and the order parameter \( \Delta  \) unchanged in first order
in magnetic field: 
\begin{equation}
\Sigma _{2}^{(1)}=0,\qquad \Delta ^{(1)}=0.
\end{equation}
 Mathematically, this is due to the fact that self-consistent solution of (\ref{Sigma1}),
(\ref{Ghat1}) produces (linear) homogeneous equations on \( \Sigma _{2}^{(1)} \)
and \( \Delta ^{(1)} \) which have regularly only trivial solutions. At the
same time, the equation on the normal part of self-energy \( \Sigma _{1}^{(1)} \)
involves a right-hand side proportional to the magnetic field coming from 
\begin{equation}
\label{H1}
\widehat{H}^{(1)}(\hat{k})=\left( \begin{array}{cc}
-\mu \f \sigma _{\alpha \beta }\mathbf{B} & \Delta _{\hat{k}\alpha \beta }^{(1)}\\
\Delta _{\hat{k}\alpha \beta }^{(1)+} & \mu \f \sigma _{\alpha \beta }^{\mathrm{T}}\mathbf{B}
\end{array}\right) .
\end{equation}

To be more specific, when written out in components, Eq. (\ref{Ghat1}) reads
\begin{eqnarray}
G^{(1)} & = & G^{(0)}\left[ 1\right] +F^{(0)}\left[ 2\right] ,\label{G1} \\
F^{(1)+} & = & F^{(0)+}\left[ 1\right] +\overline{G}^{(0)}\left[ 2\right] ,\label{F+1} 
\end{eqnarray}
 where we made use of the abbreviated notations 
\begin{eqnarray}
\left[ 1\right]  & = & (\Sigma _{1}^{(1)}-\mu \f \sigma \mathbf{B})G^{(0)}+(\Sigma _{2}^{(1)}-\Delta ^{(1)})F^{(0)+},\\
\left[ 2\right]  & = & (\Sigma _{2}^{(1)+}-\Delta ^{(1)+})G^{(0)}+(\overline{\Sigma }_{1}^{(1)}+\mu \f \sigma ^{\mathrm{T}}\mathbf{B})F^{(0)+}.
\end{eqnarray}

Let us first look on the equation on \( \Delta ^{(1)} \). Expanding Eq. (\ref{D})
yields 
\begin{equation}
\label{D1}
\Delta _{\hat{k}}^{(1)+}=\frac{3}{2}V_{1}g_{\mu }^{+}T\sum _{n}\sum _{\mathbf{k}^{\prime }}(\hat{k}\hat{k}^{\prime })\tr [g_{\mu }F^{(1)+}(\omega _{n},\mathbf{k}^{\prime })].
\end{equation}
 Substituting here (\ref{F+1}) for \( F^{(1)+} \) one obtains the equation
on \( \Delta ^{(1)+} \). It turns out that after taking trace over spin indices
as well as integrating over \( \xi  \) on the supposition of electron-hole
symmetry there remain only terms proportional to \( \Delta ^{(1)+} \).

Indeed, consider the terms in (\ref{F+1}) in pairs. The two terms involving
the anomalous part \( \Sigma _{2}^{(1)}(\omega _{n}) \) of the self-energy,
which, we remind, does not depend on momenta, are proportional to either \( \overline{G}^{(0)}G^{(0)} \),
which does not depend on \( \hat{k} \), or to \( (F^{(0)+})^{2} \) \( \propto \hat{k}^{2} \).
They vanish after averaging with \( \hat{k}^{\prime } \) over the direction
of momenta in (\ref{D1}).

The two terms proportional to \( \mathbf{B} \) when inserted into (\ref{D1})
give 
\begin{eqnarray}
 & \propto  & \tr [g_{\mu }\left( F^{(0)+}\f \sigma \mathbf{B}G^{(0)}-\overline{G}^{(0)}\f \sigma ^{\mathrm{T}}\mathbf{B}F^{(0)+}\right) ]\nonumber \\
 & \propto  & d_{\nu }\tr [g_{\mu }\left( g_{\nu }^{+}(i\widetilde{\omega }_{n}+\xi ^{\prime })\sigma _{\lambda }-\sigma _{\lambda }^{\mathrm{T}}(i\widetilde{\omega }_{n}-\xi ^{\prime })g_{\nu }^{+}\right) ]B_{\lambda }.\nonumber 
\end{eqnarray}
 The terms proportional to \( \xi ^{\prime } \) vanish on integrating over
\( \xi ^{\prime } \) in the supposition of particle-hole symmetry. The terms
proportional to \( \widetilde{\omega }_{n} \) vanish on summation over the
Matsubara frequencies \( \omega _{n}=(2n+1)\pi T \) from \( n=-\infty  \)
to \( \infty  \) because of the oddness of \( \widetilde{\omega }_{n}(\omega _{n}) \):
\( \widetilde{\omega }_{n}(-\omega _{n})=-\widetilde{\omega }_{n}(\omega _{n}) \).

Similar argumentation leads to the conclusion of vanishing of the terms involving
\( \Sigma _{1}^{(1)} \) because, as we shall see later, in the assumption of
particle-hole symmetry \( \Sigma _{1}^{(1)} \) is even in \( \omega _{n} \).

So we are left with a homogeneous linear equation on \( \Delta ^{(1)} \) which
has normally only trivial solutions \( \Delta ^{(1)}=0 \).

One obtains the equations to determine the normal \( \Sigma _{1}^{(1)} \) and
anomalous \( \Sigma _{2}^{(1)} \) parts of the self-energy by plugging (\ref{G0})
and (\ref{Sigma0}) into (\ref{Sigma1}) 
\begin{eqnarray}
\widehat{\Sigma }^{(1)} & = & \left( \begin{array}{cc}
\Sigma _{1}^{(1)} & \Sigma _{2}^{(1)}\\
\Sigma _{2}^{(1)+} & \overline{\Sigma }_{1}^{(1)}
\end{array}\right) =\frac{n_{\mathrm{imp}}u^{2}}{(1-u^{2}g^{2})^{2}}\label{Sigma1gen} \\
 & \times  & \left( \begin{array}{cc}
(1+ug)^{2}\sum _{\mathbf{k}}G^{(1)} & (1-u^{2}g^{2})\sum _{\mathbf{k}}F^{(1)}\\
(1-u^{2}g^{2})\sum _{\mathbf{k}}F^{(1)+} & (1-ug)^{2}\sum _{\mathbf{k}}\overline{G}^{(1)}
\end{array}\right) .\nonumber 
\end{eqnarray}

Since \( \Delta _{\hat{k}}^{(0)} \), and consequently also \( F^{(0)}(\omega _{n},\mathbf{k}) \)
by virtue of (\ref{G0}), and \( \Delta _{\hat{k}}^{(1)} \) are all proportional
to the unit vector \( \hat{k} \), any of the terms in (\ref{G1}), (\ref{F+1})
comprising one or three of either of the quantities vanishes on summing over
\( \mathbf{k} \) in (\ref{Sigma1}). For the anomalous part there remains only
two terms giving a homogeneous equation on the four components of \( \Sigma _{2}^{(1)} \):
\begin{equation}
\Sigma _{2}^{(1)+}=\frac{n_{\mathrm{imp}}u^{2}}{1-u^{2}g^{2}}\sum _{\mathbf{k}}\left( F^{(0)+}\Sigma _{2}^{(1)}F^{(0)+}+\overline{G}^{(0)}\Sigma _{2}^{(1)+}G^{(0)}\right) .
\end{equation}
Unless in exceptional cases, this equation has only trivial solution \( \Sigma _{2}^{(1)}\equiv 0 \).
However, one can verify separately for each component of the expansion of \( \Sigma _{2}^{(1)}(\omega _{n}) \)
in terms of the three Pauli matrices and the unit matrix:
\begin{equation}
\label{Sigma1exp}
\Sigma _{2\alpha \beta }^{(1)}(\omega _{n})=\Sigma _{2}^{(1)}(\omega _{n})\delta _{\alpha \beta }+\f \Sigma _{2}^{(1)}(\omega _{n})\f \sigma _{\alpha \beta },
\end{equation}
that this homogeneous equation on \( \Sigma _{2}^{(1)}(\omega _{n}) \) does
indeed have only trivial solution.

Taking into account the remark after Eq. (\ref{xik}) we omit the terms proportional
to \( \widehat{\tau }_{3} \) in (\ref{Sigma1gen}) and for the normal part
\( \Sigma _{1}^{(1)} \) get the equation
\begin{eqnarray}
\Sigma _{1}^{(1)} & = & n_{\mathrm{imp}}u^{2}\frac{1+u^{2}g^{2}}{(1-u^{2}g^{2})^{2}}\sum _{\mathbf{k}}[G^{(0)}(\Sigma _{1}^{(1)}-\mu \f \sigma \mathbf{B})G^{(0)}\nonumber \\
 & + & F^{(0)}(\overline{\Sigma }_{1}^{(1)}+\mu \f \sigma ^{\mathrm{T}}\mathbf{B})F^{(0)+}].\label{Sigma1calc} 
\end{eqnarray}

Making use of the explicit form of the zero-order Green function (\ref{G0}),
we first calculate the sum over \( \mathbf{k} \), introducing the auxiliary
quantity
\begin{equation}
\Pi (\omega _{n})=\frac{\pi N_{0}\Delta ^{2}}{3(\widetilde{\omega }_{n}^{2}+\Delta ^{2})^{3/2}},
\end{equation}
 so that \( \sum _{\mathbf{k}}(G^{(0)})^{2}=\frac{3}{2}\Pi (\omega _{n}) \)
and \( \sum _{\mathbf{k}}F^{(0)}F^{(0)+}=\Pi (\omega _{n}) \). In addition,
we introduce
\begin{eqnarray}
\widetilde{\Pi }(\omega _{n}) & = & n_{\mathrm{imp}}u^{2}\frac{1+u^{2}g^{2}}{(1-u^{2}g^{2})^{2}}\Pi (\omega _{n})\nonumber \\
 & = & \frac{\Gamma \Delta ^{2}}{3}\frac{\widetilde{\omega }_{n}^{2}\cos 2\delta _{0}+\Delta ^{2}\cos ^{2}\delta _{0}}{\sqrt{\widetilde{\omega }_{n}^{2}+\Delta ^{2}}\left( \widetilde{\omega }_{n}^{2}+\Delta ^{2}\cos ^{2}\delta _{0}\right) ^{2}}.
\end{eqnarray}
The second terms in (\ref{Sigma1calc}) is proportional to the product of \( F_{\alpha \beta }^{(0)}\propto \Delta \mathbf{g}_{\alpha \beta }\overleftrightarrow {R}\hat{k} \)
and \( F^{(0)+}\propto \Delta \mathbf{g}_{\alpha \beta }^{+}\overleftrightarrow {R}\hat{k} \).
After averaging over \( \hat{k} \) and integration over \( \xi  \) we find
that 
\begin{equation}
\sum _{\mathbf{k}}F^{(0)}f(\omega _{n})F^{(0)+}=\frac{1}{2}\Pi (\omega _{n})\mathbf{g}f(\omega _{n})\mathbf{g}^{+}.
\end{equation}
So
\begin{eqnarray}
\Sigma _{1}^{(1)} & = & \frac{3}{2}\widetilde{\Pi }(\omega _{n})\left[ \Sigma _{1}^{(1)}-\mu \f \sigma \mathbf{B}+\frac{1}{3}\mathbf{g}(\overline{\Sigma }_{1}^{(1)}+\mu \f \sigma ^{\mathrm{T}}\mathbf{B})\mathbf{g}^{+}\right] ,\label{Xi1} \\
\overline{\Sigma }_{1}^{(1)} & = & \frac{3}{2}\widetilde{\Pi }(\omega _{n})\left[ \overline{\Sigma }_{1}^{(1)}+\mu \f \sigma ^{\mathrm{T}}\mathbf{B}+\frac{1}{3}\mathbf{g}^{+}(\Sigma _{1}^{(1)}-\mu \f \sigma \mathbf{B})\mathbf{g}\right] ,\label{Xi1over} 
\end{eqnarray}

Expanding \( \Sigma _{1}^{(1)} \) and \( \overline{\Sigma }_{1}^{(1)} \) in
the basis of the three Pauli matrices and the unit matrix similarly to (\ref{Sigma1exp}),
one sees that for the coefficient of expansion over the unit matrix one obtains
an uncoupled homogeneous equation, with trivial solution. So one may put simply
\begin{equation}
\Sigma _{1}^{(1)}=\f \Sigma _{1}^{(1)}\f \sigma ,\qquad \overline{\Sigma }_{1}^{(1)}=\overline{\f \Sigma }_{1}^{(1)}\f \sigma ^{\mathrm{T}},
\end{equation}

Using \( \mathbf{g}=i\f \sigma \sigma _{y} \), \( \mathbf{g}^{+}=-i\sigma _{y}\f \sigma  \)
and the equalities 
\begin{equation}
\label{sigmamu}
\sigma _{y}\f \sigma \sigma _{y}=-\f \sigma ^{\mathrm{T}},\qquad \sigma _{\mu }\f \sigma \sigma _{\mu }=-\f \sigma 
\end{equation}
we find that 
\begin{equation}
\label{g}
g_{\mu }\f \sigma g^{+}_{\mu }=g^{+}_{\mu }\f \sigma g_{\mu }=\f \sigma ^{\mathrm{T}}.
\end{equation}
So that solution of the system (\ref{Xi1}), (\ref{Xi1over}) is 
\begin{equation}
\Sigma _{1}^{(1)}=-\mu \f \sigma \mathbf{B}\frac{\widetilde{\Pi }(\omega _{n})}{1-\widetilde{\Pi }(\omega _{n})},\qquad \overline{\Sigma }_{1}^{(1)}=\mu \f \sigma ^{\mathrm{T}}\mathbf{B}\frac{\widetilde{\Pi }(\omega _{n})}{1-\widetilde{\Pi }(\omega _{n})}.
\end{equation}
Since from (\ref{G1}) it follows that
\begin{equation}
G^{(1)}=G^{(0)}(\Sigma _{1}^{(1)}-\mu \f \sigma \mathbf{B})G^{(0)}+F^{(0)}(\overline{\Sigma }_{1}^{(1)}+\mu \f \sigma ^{\mathrm{T}}\mathbf{B})F^{(0)+},
\end{equation}
we find from (\ref{M}) that
\begin{equation}
\mathbf{M}=-2\mu ^{2}\mathbf{B}T\sum _{n}\frac{\Pi (\omega _{n})}{1-\widetilde{\Pi }(\omega _{n})}.
\end{equation}

Because we have first integrated over \( \mathbf{k} \) and left the summation
over \( n \) this expression does not include \cite{Abrikosov63} normal state
Fermi gas Pauli susceptibility \( \chi ^{0}_{n}=2\mu ^{2}N_{0} \). Adding and
subtracting it from both sides of (\ref{M}), we get eventually
\begin{equation}
\frac{\chi ^{0}}{\chi ^{0}_{n}}=1-\frac{\pi }{3}T\sum _{n}\frac{\DS \frac{\Delta ^{2}}{(\widetilde{\omega }_{n}^{2}+\Delta ^{2})^{3/2}}}{\DS 1-\frac{\Gamma \Delta ^{2}}{3}\frac{\widetilde{\omega }_{n}^{2}\cos 2\delta _{0}+\Delta ^{2}\cos ^{2}\delta _{0}}{\sqrt{\widetilde{\omega }_{n}^{2}+\Delta ^{2}}\left( \widetilde{\omega }_{n}^{2}+\Delta ^{2}\cos ^{2}\delta _{0}\right) ^{2}}}.
\end{equation}
This expression gives susceptibility as a function of temperature for a given
scattering rate \( \Gamma  \) and a scattering phase \( \delta _{0} \). The
infinite sum over the Matsubara frequencies should be taken using self-consistent
expressions for the renormalized Matsubara frequency \( \widetilde{\omega }_{n} \)
(\ref{omegaTilde2}) and the order parameter \( \Delta (T) \) (\ref{logTTc0}).

In the clean limit \( \Gamma \to 0 \) the Matsubara frequencies rest unrenormalized
and susceptibility reduces to its conventional bulk value
\begin{equation}
\frac{\chi ^{0}}{\chi ^{0}_{n}}=1-\frac{\pi }{3}T\sum _{n}\frac{\Delta ^{2}}{(\omega _{n}^{2}+\Delta ^{2})^{3/2}}\equiv 1-\frac{1}{3}\left( 1-Y_{B}(T)\right) ,
\end{equation}
 where 
\begin{equation}
Y_{B}(T)=1-\pi T\sum _{n}\frac{\Delta ^{2}}{(\omega _{n}^{2}+\Delta ^{2})^{3/2}}
\end{equation}
 is the B-phase Yosida function.

At \( T\to T_{c} \) susceptibility \( \chi ^{0}/\chi ^{0}_{n} \) behaves linearly.
Indeed, then \( \Delta \to 0 \) and \( \widetilde{\omega }\to \omega +\Gamma  \).
Hence
\begin{equation}
\frac{\chi ^{0}}{\chi ^{0}_{n}}\to 1-\frac{\pi \Delta ^{2}}{3}T_{c}\sum _{n}\frac{1}{|\widetilde{\omega }_{n}|^{3}}\to 1+\frac{2\Delta ^{2}}{3(4\pi T_{c})^{2}}\psi ^{(2)}_{c},
\end{equation}
where \( x_{c}=\Gamma /2\pi T_{c} \) and \( \psi _{c}^{(n)}=\psi ^{(n)}(\frac{1}{2}+x_{c}) \)
is the polygamma function. Using the asymptotic expression (\ref{DeltaAtTtoTc})
for \( \Delta  \) at \( T\to T_{c} \), we find eventually

\begin{equation}
\label{XXn}
\left. \frac{\chi ^{0}}{\chi ^{0}_{n}}\right| _{T\to T_{c}}\to 1+2\psi _{c}^{(2)}\frac{x_{c}\psi _{c}^{(1)}-1}{3\psi _{c}^{(2)}+x_{c}\psi _{c}^{(3)}\cos 2\delta _{0}}\frac{T_{c}-T}{T_{c}}.
\end{equation}

In the Born limit (\( \delta _{0}\to 0 \)) we arrive at the expression for
the spin susceptibility
\begin{equation}
\label{ChiBorn}
\frac{\chi ^{0}}{\chi ^{0}_{n}}=1-\frac{\pi }{3}T\sum _{n}\frac{\DS \frac{\Delta ^{2}}{(\widetilde{\omega }_{n}^{2}+\Delta ^{2})^{3/2}}}{\DS 1-\frac{\Gamma \Delta ^{2}}{3}\frac{1}{(\widetilde{\omega }_{n}^{2}+\Delta ^{2})^{3/2}}}.
\end{equation}
In the opposite unitary limit (\( \delta _{0}\to \frac{\pi }{2} \))
\begin{equation}
\label{ChiUnitary}
\frac{\chi ^{0}}{\chi ^{0}_{n}}=1-\frac{\pi }{3}T\sum _{n}\frac{\DS \frac{\Delta ^{2}}{(\widetilde{\omega }_{n}^{2}+\Delta ^{2})^{3/2}}}{\DS 1+\frac{\Gamma \Delta ^{2}}{3}\frac{1}{\widetilde{\omega }_{n}^{2}\sqrt{\widetilde{\omega }_{n}^{2}+\Delta ^{2}}}}.
\end{equation}

Susceptibilities for several impurity concentrations and three scattering phases
(\( \delta _{0}=0 \) --- Born limit, \( \delta _{0}=\frac{\pi }{4} \), and
\( \delta _{0}=\frac{\pi }{2} \) --- unitary limit) are plotted as functions
of the reduced temperature \( T/T_{c} \) on Fig. \ref{Fig-severalX} with the
Fermi liquid effects taken into account (see below). 
\begin{figure}[!t]
{\par\centering \resizebox*{1\columnwidth}{!}{\includegraphics{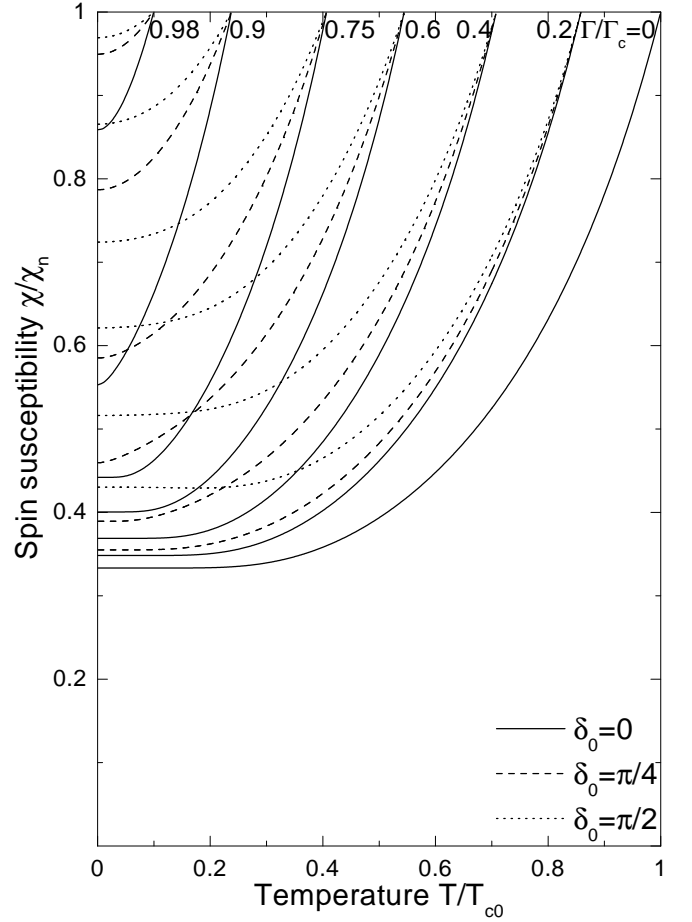}} \par}

\caption{\label{Fig-severalX} Spin susceptibility of superfluid \protect\( ^{3}\protect \)He
in aerogel versus temperature for several values of impurity concentration \protect\( \Gamma \protect \)
and scattering phases \protect\( \delta _{0}\protect \) with the Fermi liquid
corrections taken into account. Solid lines correspond to the Born limit \protect\( \delta _{0}=0\protect \),
dashed lines to \protect\( \delta _{0}=\frac{\pi }{4}\protect \) and dotted
lines to the unitary scattering \protect\( \delta _{0}=\frac{\pi }{2}\protect \).
In order of the decrease of the superfluid transition temperature \protect\( T_{c}\protect \)
the triples of the curves correspond to \protect\( \Gamma /\Gamma _{c}=\protect \)0,
0.2, 0.4, 0.6, 0.75, 0.9, 0.98. The first curve in the pure case \protect\( \Gamma =0\protect \)
is common for all phases.}
\end{figure}

\subsection{Fermi-liquid corrections}

Fermi-liquid interaction between quasiparticles leads to renormalization of
the susceptibility. An external magnetic field is screened by a polarization
of the liquid, which is quantitatively described by even terms of the expansion
in Legendre polynomials of the antisymmetric (exchange) part \( F^{\mathrm{a}} \)
of the Fermi-liquid interaction. 

It was argued in Ref. \cite{Thuneberg01} on the basis of experimental data
that only the zeroth term \( F_{0}^{\mathrm{a}} \) is non-zero. Then the magnetization
\( M_{\alpha }=\chi _{\alpha \beta }B_{\beta } \) induced by magnetic field
\( \mathbf{B} \) in a liquid with interaction equals the magnetization \( M_{\alpha }=\chi _{\alpha \beta }^{0}B_{\beta }^{\mathrm{eff}} \)
that would be induced in an interaction-free liquid by an effective field
\begin{equation}
\mathbf{B}^{\mathrm{eff}}=\mathbf{B}-F_{0}^{\mathrm{a}}\mathbf{M}/\chi _{n}^{0}.
\end{equation}
Here \( \chi _{n}^{0}=2\mu ^{2}N_{0} \) is the Pauli paramagnetic susceptibility
of a normal Fermi-gas. This yields \cite{Leggett65}, \cite{Vollhardt}
\begin{equation}
\chi _{\alpha \beta }=\left( \delta _{\alpha \beta }+F_{0}^{\mathrm{a}}\chi _{\alpha \beta }^{0}/\chi _{n}^{0}\right) ^{-1}\chi _{\alpha \beta }^{0}.
\end{equation}
 In the B-phase \( \chi _{\alpha \beta }^{0}\propto \delta _{\alpha \beta } \)
and we get 
\begin{equation}
\label{ChiChin}
\frac{\chi }{\chi _{n}}=\frac{1+F_{0}^{\mathrm{a}}}{1+F_{0}^{\mathrm{a}}\frac{\chi ^{0}}{\chi _{n}^{0}}}\frac{\chi ^{0}}{\chi _{n}^{0}},
\end{equation}
 where \( \chi _{n}=\chi _{n}^{0}/(1+F_{0}^{\mathrm{a}}) \) is the Fermi-liquid
susceptibility. In \( ^{3} \)He \( F_{0}^{\mathrm{a}}\approx -\frac{3}{4} \)
slightly varying with pressure.

For the asymptotics at \( T\rightarrow T_{c} \) we have 
\begin{equation}
\left. \frac{\chi }{\chi _{n}}\right| _{T\rightarrow T_{c}}=1+\left. \frac{d(\chi /\chi _{n})}{d(T/T_{c})}\right| _{T\rightarrow T_{c}}\frac{T-T_{c}}{T_{c}},
\end{equation}
 where from (\ref{ChiChin}) 
\begin{equation}
\left. \frac{d(\chi /\chi _{n})}{d(T/T_{c})}\right| _{T\rightarrow T_{c}}=\frac{1}{1+F_{0}^{\mathrm{a}}}\left. \frac{d(\chi ^{0}/\chi ^{0}_{n})}{d(T/T_{c})}\right| _{T\rightarrow T_{c}}
\end{equation}
 is the slope of the plots of \( \chi  \) versus \( T \) at \( T\to T_{c} \).

Taking \( -\frac{3}{4} \) as a value for \( F_{0}^{\mathrm{a}} \), we see
that the slope of the curve \( \chi \left( T\right)  \) at \( T=T_{c} \) with
the Fermi-liquid interaction taken into account, is 
\begin{equation}
\label{coeffprop}
1/(1+F_{0}^{\mathrm{a}})\approx 4
\end{equation}
 times greater than that of \( \chi ^{0}(T) \) --- without the interaction. 

\begin{figure}[!t]
{\par\centering \resizebox*{1\columnwidth}{!}{\includegraphics{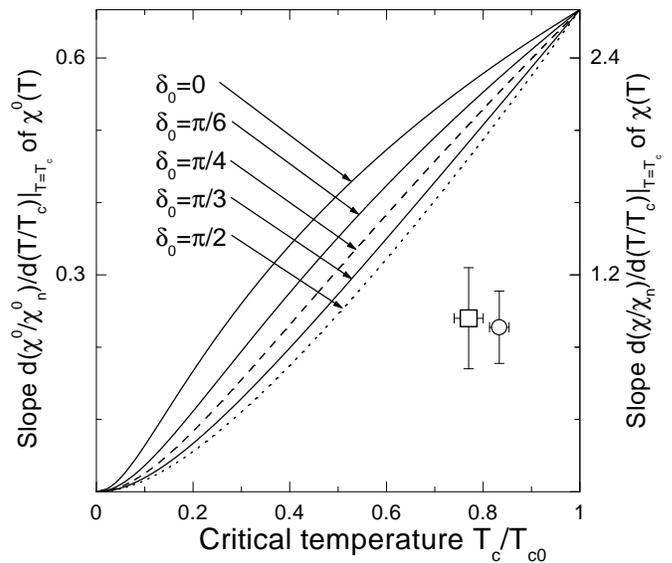}} \par}

\caption{\label{Fig-slopes} The slope at \protect\( T\to T_{c}\protect \) of the reduces
temperature \protect\( T/T_{c}\protect \) plots of the ``bare'' susceptibility
\protect\( \chi ^{0}/\chi _{n}^{0}\protect \) (left scale) or the susceptibility
\protect\( \chi /\chi _{n}\protect \) with the Fermi liquid corrections (right
scale, coefficient of proportionality (\ref{coeffprop})) as functions of the
suppression \protect\( T_{c}/T_{c0}\protect \) of the critical temperature.
The crosses mark the experimental values of the slopes in \cite{sprague96}
(\protect\( \square \protect \)) and in \cite{Osheroff00} (\protect\( \bigcirc \protect \)).}
\end{figure}

The slopes of both \( \chi ^{0}\left( T\right)  \) and \( \chi \left( T\right)  \)
for \( F_{0}^{\mathrm{a}}=-\frac{3}{4} \) versus the suppression \( T_{c}/T_{c0} \)
of the critical temperature are plotted on Fig. \ref{Fig-slopes} in two scales
of the ordinate axis --- the left one for the slope of \( \chi ^{0}\left( T\right)  \),
and the right one for the slope of \( \chi \left( T\right)  \), which differs
from the former in (\ref{coeffprop}) times. On the same picture we also pointed
the slopes experimentally observed in \cite{sprague96} at 18.7 bars and in
\cite{Osheroff00} at 32 bars --- the only experimental data present up to day
the authors are aware of. The error bars span the possible slopes of the linear
fits to the data at \( T\lesssim T_{c} \) within the plotted error bars of
the respective experimental works.

\section{Conclusions}

In the framework of the homogeneous scattering model we found spin susceptibility
of \( ^{3} \)He in aerogel considering the latter as ordinary randomly distributed
impurities with an arbitrary scattering phase \( \delta _{0} \) in \( s \)-wave
channel. The answer is given in the form of an infinite sum over the Matsubara
frequencies \( \omega _{n}=(2n+1)\pi T \) with the summation term depending
on the amplitude \( \Delta  \) of the order parameter and the self-consistently
renormalized Matsubara frequency \( \widetilde{\omega }_{n} \). Numerically
plotted curves \( \chi (T) \) show that susceptibility is less suppressed in
comparison to the bulk dependence, this suppression being less pronounced for
larger \( \delta _{0} \). 

Comparison with high-pressure experimental data (18.7 bars \cite{sprague96}
and 32 bars \cite{Osheroff00}) shows that the slope of \( \chi (T) \) at \( T\to T_{c} \)
is approximately twice less than the value predicted by the theory. The discrepancy
is obviously unavoidable because of the major inadequacy of the homogeneous
scattering model at high pressures, where the superfluid coherence length becomes
shorter than the correlation length of the internal structure of aerogel \cite{porto99}.
Note that quantitatively the difference between experimental and theoretical
values seems to be in line with the fact that experimentally observed at high
pressures (19.7 bars) superfluid density \cite{Alles98} was approximately by
the factor of 2 smaller than predicted by an Abrikosov-Gor'kov theory \cite{Hanninen00}. 

On the other hand, we pointed out that for any finite scattering phase \( \delta _{0}>0 \)
the impurity concentration threshold of the gapless superfluidity onset diminishes
dramatically. For example, for the value \( \delta _{0}=\frac{\pi }{4} \) which
was maintained to be the most satisfactory phase for the real structure of aerogel,
gapless superfluidity sets on in the whole temperature range starting at the
critical temperature suppression of \( T_{c}\approx 0.51T_{c0} \) compared
to \( T_{c}\approx 0.22T_{c0} \) in the Born limit \( \delta _{0}\to 0 \).
Whereas in the unitary limit \( \delta _{0}\to \frac{\pi }{2} \) the superfluid
is always gapless. For impurity concentrations exceeding the gapless threshold
all the thermodynamical quantities vanish algebraically at zero temperature.
Experimental observation of such behavior could provide additional immediate
insight into the quasiparticle scattering on impurities in aerogel.

\section{Acknowledgments}

We wish to thank J. A. Sauls for valuable comments on the manuscript and attraction
of our attention to the preprint \cite{Priya01}, which complies with ours in
the main results. One of the authors (P. K.) is grateful to J. Flouquet for
the granted opportunity to complete the work at the Center for Nuclear Research
in Grenoble.

\appendix

\section{Asymptotical behavior of the order parameter \protect\( \Delta \protect \)
at \protect\( T\to T_{c}\protect \) and \protect\( T\to 0\protect \) }

In this Appendix we elaborate on the details of the calculation of the asymptotics
of the order parameter already utilized in the main text of the paper.

\subsection{Asymptotics at the critical temperature}

Expanding (\ref{omegaTilde2}) in the degrees of \( \Delta ^{2} \) gives (we
consider only positive \( \omega _{n} \) because of the oddness of \( \widetilde{\omega }_{n}(\omega _{n}) \):
\( \widetilde{\omega }_{n}(-\omega _{n})=-\widetilde{\omega }_{n}(\omega _{n}) \))
\begin{equation}
\widetilde{\omega }_{n}\approx \omega _{n}+\Gamma +\Gamma \frac{1-2\cos ^{2}\delta _{0}}{2(\omega _{n}+\Gamma )}\Delta ^{2}+O(\Delta ^{4}).
\end{equation}

When inserted into (\ref{logTTc0}) it yields in first order in \( \Delta ^{2} \)
the asymptotics at \( T\rightarrow T_{c} \)
\begin{equation}
\label{DeltaAtTtoTc}
\Delta ^{2}=3(4\pi T_{c})^{2}\frac{x_{c}\psi _{c}^{(1)}-1}{3\psi _{c}^{(2)}+x_{c}\psi _{c}^{(3)}\cos 2\delta _{0}}\frac{T_{c}-T}{T_{c}}.
\end{equation}
 Here \( x_{c}=\Gamma /2\pi T_{c} \) and \( \psi _{c}^{(n)}=\psi ^{(n)}(\frac{1}{2}+x_{c}) \),
where 
\begin{equation}
\psi ^{(n)}(z)=\partial _{z}^{n}\psi (z)=\sum _{k=0}^{\infty }\frac{(-1)^{n+1}n!}{(z+k)^{n+1}}
\end{equation}
 is the polygamma function. At \( \Gamma \rightarrow 0 \) also \( x_{c}\rightarrow 0 \)
and \( \psi _{c}^{(2)}\rightarrow \psi ^{(2)}(\frac{1}{2})=-14\zeta (3) \),
and thus (\ref{DeltaAtTtoTc}) transforms into the conventional BCS asymptotics
\begin{equation}
\Delta ^{2}=\frac{8\pi ^{2}T^{2}_{c0}}{7\zeta (3)}\frac{T_{c0}-T}{T_{c0}}.
\end{equation}

\subsection{Order parameter at zero temperature}

Consider the sum \( \pi T\sum _{n}(\omega _{n}^{2}+\Delta ^{2})^{-1/2} \).
When \( T\rightarrow 0 \) the summation over the Matsubara frequencies \( \omega _{n} \)
may be substituted with an integration over the continuous variable \( \omega  \)
\begin{equation}
\label{LogDelta}
\pi T\sum _{n}\frac{1}{\sqrt{\omega _{n}^{2}+\Delta ^{2}}}\overset {T\rightarrow 0}{\longrightarrow }\int _{0}^{\epsilon _{1}}\frac{d\omega }{\sqrt{\omega ^{2}+\Delta ^{2}}}\approx \log \frac{2\epsilon _{1}}{\Delta },
\end{equation}
 where the divergent integration up to infinity is cut off at \( \epsilon _{1} \). 

The general temperature dependence of the order parameter is obtained by combination
of (\ref{1/NV}), (\ref{Delta00}) and (\ref{LogDelta})
\begin{equation}
\label{LogDeltaDelta00}
\log \frac{\Delta }{\Delta _{00}}=\pi T\sum _{n}\left( \frac{1}{\sqrt{\widetilde{\omega }_{n}^{2}+\Delta ^{2}}}-\frac{1}{\sqrt{\omega _{n}^{2}+\Delta ^{2}}}\right) .
\end{equation}

At \( T\rightarrow 0 \) the value of the order parameter tends to the value
\( \Delta _{0} \) the implicit dependence of which on \( \Gamma  \) is set
by replacing summation in the above expression with an integral. The integral
over \( \omega  \) from the first term in the parentheses in (\ref{LogDeltaDelta00})
transforms into an integral over \( \widetilde{\omega } \) by means of the
substitution (\ref{omegaTilde2}) and eventually we get an equation that determines
implicitly the impurity concentration behavior of the value of the order parameter
at zero temperature:
\begin{equation}
\label{implOnD0}
\frac{1}{\Gamma }\log \frac{\Delta _{0}}{\Delta _{00}}+\frac{\pi }{2\Delta _{0}(1+\cos \delta _{0})}=0,
\end{equation}
if \( \Gamma <\Delta _{0}\cos ^{2}\delta  \) and 
\begin{eqnarray}
0 & = & \frac{1}{\Gamma }\log \frac{\widetilde{\omega }_{0}+\sqrt{\widetilde{\omega }_{0}^{2}+\Delta _{0}^{2}}}{\Delta _{00}}+\frac{\pi }{2\Delta _{0}(1+\cos \delta _{0})}\label{implOnD0-1} \\
 & - & \frac{\widetilde{\omega }_{0}}{\widetilde{\omega }_{0}^{2}+\Delta _{0}^{2}\cos ^{2}\delta _{0}}\nonumber \\
 & - & \frac{1}{\Delta _{0}\sin ^{2}\delta _{0}}\left( \cos \delta _{0}\arctan \frac{\widetilde{\omega }_{0}}{\Delta _{0}\cos \delta _{0}}-\arctan \frac{\widetilde{\omega }_{0}}{\Delta _{0}}\right) \nonumber 
\end{eqnarray}
 if \( \Gamma >\Delta _{0}\cos ^{2}\delta  \). Here 
\begin{equation}
\widetilde{\omega }_{0}(\Gamma ,\Delta )=\sqrt{\frac{\Gamma ^{2}-2\Delta ^{2}\cos ^{2}\delta _{0}+\Gamma \sqrt{\Gamma ^{2}+4\Delta ^{2}\sin ^{2}\delta _{0}}}{2}}
\end{equation}
 is the solution of (\ref{omegaTilde2}) for \( \omega _{n}=0 \) and in (\ref{implOnD0-1})
it was taken with the arguments \( \Gamma ,\Delta _{0} \).

For small \( \Gamma  \) Eq. (\ref{implOnD0}) gives 
\begin{equation}
\frac{\Delta _{0}}{\Delta _{00}}\approx 1-\frac{\pi }{4(1+\cos \delta _{0})}\frac{\Gamma }{\Gamma _{c}},
\end{equation}
 while for \( \Gamma  \) close to the critical value 
\begin{equation}
\frac{\Delta _{0}}{\Delta _{00}}\approx \sqrt{\frac{1}{1-\frac{2}{3}\cos 2\delta _{0}}\frac{\Gamma _{c}-\Gamma }{\Gamma _{c}}}.
\end{equation}

\end{document}